\documentstyle[11pt,newpasp,twoside,epsf]{article}
\def\edcomment#1{\iffalse\marginpar{\raggedright\sl#1\/}\else\relax\fi}
\reversemarginpar
\begin{document}
\title{Tests of a Structure for Quasars} 
\author{Martin Elvis}
\affil{Harvard-Smithsonian Center for Astrophysics,
60 Garden St.,  Cambridge MA02138 USA}

\begin{abstract}
The model I recently proposed for the structure of quasars
unifies all the emission, absorption and reflection
phenomenology of quasars and AGN, and so is heavily
overconstrained and readily tested.

Here I concentrate on how the model has performed against tests
since publication - with many of the tests being reported at this
meeting. I then begin to explore how these and future tests can
discriminate between this wind model and 3 well-defined
alternatives.

\end{abstract}

\section{Introduction}

Quasar research suffers from an overabundance of phenomenology.
There has been a constant piling up of new details at all
wavelengths with sadly little integration, let alone physical
explanation. While this is fascinating for insiders,
quasars have become an intimidating field for the general
astronomer. 

This is not a situation to despair of, because stars were in a
similar situation for 20 years (Lawrence 1987). The spectroscopic
definitions of the stellar types (O~B~A~F~G~K~M) read quite
similarly to those of AGN classifications (e.g. G stars: ``CaII
strong; Fe and other metals strong; H weaker'', Allen 1975).  We
now know that the main sequence is a simple temperature
progression, determined fundamentally by stellar mass. There is
hope that the complexities of quasars will resolve themselves the
same way.  I have proposed (Elvis 2000) a geometrical and
kinematic model for quasars that does seem to subsume a great
deal of the phenomenology of quasar emission and absorption lines
into a simple scheme.

Quasars, unlike stars, are not spherical. (We have known that
axisymmetry is appropriate since the first double radio sources
were discovered). This means that geometry matters, and when this
is the case the physics cannot be worked out until we get the
structure right: the solar system simply could not be solved in a
Ptolemaic geometry. A normal sequence in constructing a physical
theory is to work out the right geometry, then the kinematics and
lastly the dynamics (c.f. Copernicus - Kepler - Newton). In
quasars instead, I believe that the physics has largely already
been worked out, but discarded because the geometry was not in
place, making the physics appear wrong.

Here I briefly outline the model and then concentrate on tests of
the model, many reported at this meeting. A model needs a rival
if the tests are to be strong, so I have also begun to explore
alternative wind geometries to see how they differ from my model.

\section{A Structure for Quasars}

In Elvis (2000) I proposed that a flow of warm ($\sim 10^6$K) gas
rises vertically from a {\em narrow range of radii} on an
accretion disk. This flow then accelerates, angling outward (most
likely under the influence of radiation pressure from the intense
quasar continuum) until it forms thin conical wind moving
radially (figure~1).  When the continuum source is viewed through
this wind it shows narrow absorption lines (NALs) in both UV and
X-ray (the X-ray `warm absorber'); when viewed down the radial
flow the absorption is stronger and is seen over a large range of
velocities down to $v(vertical)$, the `detachment velocity', so
forming the Broad Absorption Line quasars. Given the narrowness
of the vertical flow ($\sim 0.1r$), the divergence of the
continuum radiation at the turning point will be $\sim$6$^{\circ}$, giving 10\% solid
angle coverage, and so the correct fraction of BAL quasars.  [The
angle to the disk axis, 60$^{\circ}$, is at present arbitrarily
chosen to give the correct number of NAL and non-NAL quasars.]

\begin{figure}
\plotone{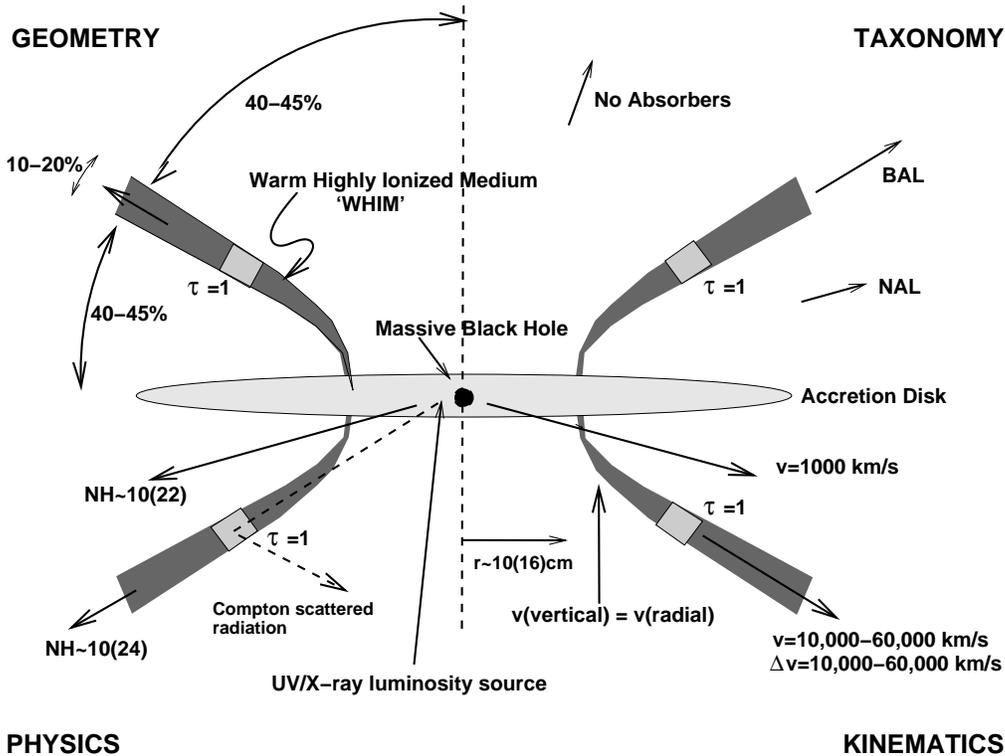}
\caption{A Structure for Quasars}
\end{figure}

This `Warm Highly Ionized Medium' (WHIM) has a cool phase (like
the ISM) with which it is in pressure equlibrium. This cool phase
provides the clouds that emit the Broad Emission Lines
(BELs). Since the BEL clouds move along with the WHIM they are
not ripped apart by shear forces; and since the medium is only
Compton thick along the radial flow direction rapid continuum
variations are not smeared out. Both problems had long been a
strong objections to pressure confined BEL clouds, but they are
invalid in the proposed geometry.

The radial flow is Compton thick along the flow direction, and so
will scatter all wavelengths passing along that direction. Since
the flow is highly non-spherical the scattered radiation will be
polarized. The solid angle covered by the radial flow is
10\%-20\%, so this fraction of all the continuum radiation will
be scattered, leading to the filling in of the BAL troughs, and
to an X-ray Compton hump in all AGN. Since the WHIM is only
ionized to FeXVII, there will be Fe-K fluorescence off the same
structure at $\sim$100~eV EW. Some of the BEL radiation will also
pass along the flow and will be scattered off the fast moving
flow, producing the polarized, non-variable `Very Broad Line
Region'.

\section{Tests of the Model}

The model makes five main claims. Each one has been tested
already and passed, and each can be tested further:

\noindent\underline{\em Claim~1: The quasar wind is a warm,
highly ionized medium (WHIM).~}
The velocities of the UV absorption line systems must match X-ray
absorbers. In NGC3783 (Kaspi, these proceedings) and NGC4051
(Collinge et al., 2001) they do match.
The line strengths must also be consistent across the X-ray and
UV lines. In NGC3783 (Kaspi, these proceedings) they also match,
if a particular EUV spectral slope is assumed. This slope is
tightly constrained, which implies an accurate thermometer for
the, otherwise hard to observe but energetically important, EUV
continuum.  Consistency with the NLR optical coronal lines is
required.
Accretion disk models predict specific EUV continuum shapes,
which will be stringently tested if the X-ray/UV absorbers are
the same.

\noindent\underline{\em Claim~2: BAL medium = NAL medium = WHIM.~}
Since Compton scattering is wavelength independent the BAL
covering factor in the soft X-ray band should be the same as in
the optical.  Partial covering of the right order is seen in the
X-ray spectrum of one BAL quasar (Mathur, these proceedings).
The low energy X-rays should be polarized like
the optical BAL troughs, and have consistent fluxes.

\noindent\underline{\em Claim~3: WHIM = BELR Confining Medium.~}
Everett (these proceedings) gave good arguments for a 2-phase
medium in quasar winds, although in a different context.
The high ionization BELs (PVII, NeVIII) at $\sim$770\AA (rest)
seen in some high redshift quasars (Hamann et al. 1995) could
arise from the WHIM. In this case they should match the WHIM
properties.  In fact they do, with column densities of
$\sim$10$^{22}$cm$^{-2}$, a temperature of 5$\times$10$^{5}$K (if
collisional), and a covering factor of 0.5 (and unlike the BELR
for which 0.1 is a typical value).
The radius of the NAL from the continuum source has to match the
radius of the high ionization BELR. In NGC~5548 this works, but
the NAL radius is poorly constrained (a factor 100). Short
recombination times, which will determine the NAL density, are
the key to a better constraint on the radius. Timescales of hours
are likely, and require quite large continuum variations on this
timescale or less to be measurable.

\noindent\underline{\em Claim~4: Narrow Range of radii:
cylindrical WHIM/BELR.~} 
Arav (these proceedings) shows that the covering factor or the
NAL systems in NGC~5548 varies systematically with velocity in a
way that is easily understood as a narrow flow of material across
our line of sight.  These constraints will determine many
parameters of the flow (thickness, angles, density contrast,
acceleration rate).

Cohen (these proceedings) explains the rotation of polarization
position angle of H$\alpha$ in some radio galaxies by means of a
rotating circularly symmetric structure, a disk or cylinder,
which emits the BELs and scatters BEL photons off the opposite
sides. A few percent of the H$\alpha$ photons are polarized and
red or blue shifted according to where they strike the opposite
face. A WHIM shaped cylinder is hollow, making it simple for
photons to cross to the other side.  Similar PA rotations are
implied for {\em all} edge-on AGN , i.e. those with NALs.

Laor (these proceedings) showed that weak EW([OIII]), the
presence of NALs, and broad CIV are correlated. If [OIII] is
isotropic and the continuum is from a disk then the EW([OIII]) is
an inclination indicator, so these effects are qualitatively
predicted by the proposed structure.

Murray (comments at this meeting) notes that following a flare
the BELs may become optically thin, revealing a double-peaked
profile for a cylinder or disk geometry. He finds just such an
effect in the NGC~5548 data base.

\noindent\underline{\em Claim~5: Reflection from Structure.~}
Compton scattering off the radial part of the structure must
produce a symmetric X-ray 6.4~keV Fe-K emission line with a width
broader but comparable to the BELR widths. Since the WHIM is
optically thin to Fe-K, while BEL clouds are optically thick, the
two profiles will be different, with Fe-K possibly showing a
double peaked structure. This needs detailed calculations, since
it can be measured with Chandra grating spectra.

The Fe-K emission line will respond to continuum changes with a
smeared response determined by the size of the $\tau$=1
scattering ring. Takashi, Inoue \& Dotani (2001) have found such
an effect in NGC~4151, and derive an Fe-K scattering region size
of 10$^{17}$cm. This is a few times larger than the CIV radius
for NGC~4151 (9$\pm$2~light-days, 2$\times$10$^{16}$cm, Kaspi et
al. 1996), consistent with our structure, but is strongly
inconsistent with an accretion disk or pc-scale torus origin

At any given angle there will be four dominant delay times
relative to a central continuum flaring event, as the flare
scatters off the near and far parts of the $\tau$=1 rings above
and below the disk plane. (The disk may obscure one or both parts
of the lower ring.) These should show up in autocorrelation
functions of the continuum at low amplitudes.  Schild (1996,
2001) has found autocorrelation timescales in the gravitational
lens Q0957+561 and shows them to be consistent with the double
ring expected from this structure.

The X-ray `Compton Hump' should show the same time smearing and
delays as the Fe-K line. Moreover the Compton Hump should be
polarized, just as the optical BAL troughs are polarized.

\section{Alternative Wind Models}

To reconcile the presence of both a fast (BAL) wind and a slow
(NAL) wind in most AGN and quasars, we adopted a single wind with
a specific geometry. It is worth considering alternative means of
reconciling these, relaxing the constraint that they are the same
flow. There are two main possibilities: either high luminosity
objects have faster winds, or faster winds are emitted in some
preferential direction , and slower winds in others. Here we take
a first look at these options. We assume that the models retain
all the other features of our model. I.e. they still try to
combine the UV and X-ray absorbers in a single WHIM (our starting
point); and they have the BEL clouds embedded in this wind; they
try to explain reflection features via scattering off the fast
wind.

\subsection{Luminosity Dependent Velocities}

In this hypothesis the fast BAL winds are emitted only by high
luminosity quasars (into a $\sim$10\% solid angle), while NAL
winds are found only at lower luminosities (for a $\sim$50\%
solid angle). The wind no longer needs to arise from a narrow
range of disk radii. There are a number of comments that can be
made:

\noindent (1) In this scenario a continuum of widths should be
found with a covering factor that decreases from $\sim$50\% at
NGC~5548 luminosities (L$_X\sim$5$\times$10$^{43}$erg~s$^{-1}$,
Elvis et al., 1978), to $\sim$10\% L$_X\sim$10$^{46}$erg~s$^{-1}$
PHL~5200, Mathur, Elvis, \& Singh 1995).  There is a range of BAL
velocities, but a line width vs. luminosity or covering factor
analysis has not yet been performed.

\noindent (2)
This model has a built-in explanation for the absence of low
luminosity BALs; our model requires low luminosity AGN to be
dustier. 

\noindent (3)
If the BELs are embedded in the fast BAL wind then they should
have similar profiles, with no obvious physical cause
for a `detachment velocity'. 

\noindent (4) Can high ionization BELs with large covering factor
arise in this model?

\noindent (5) The scattering effects of the radial part of the
wind (X-ray Fe-K emission line, Compton hump, optical polarized
flux) will disappear in lower luminosity objects since the column
density needed to produce scattering is much larger than the
column density through the X-ray warm absorbers, and low
luminosity AGN will have no fast wind out of the line of sight to
produce the polarized, non-variable `Very Broad Line Region'.
The opposite is observed (Iwasawa \& Taniguchi 1993).

\noindent (6)
The BAL opening angle of 10\% does not arise naturally from the
narrow width of the wind origin site on the disk.

\noindent (7)
Are there directions in which one looks through the wind?
If so then NALs will be seen, the wind originates from a
restricted range of radii and the picture reverts to something
close to our model.

The absence of reflection effects in this version considerably
weakens its unifying power.

\subsection{Orientation Dependent Velocities}

In this hypothesis the quasar wind has fast (BAL) velocities only
in some directions (covering $\sim$10\% solid angle), and has
slower (NAL) velocities in other directions (covering $\sim$50\%
solid angle). There are two obvious preferential directions for
the fast wind: {\em polar} and {\em equatorial}.  Some comments
are:

\noindent (1)
Proga (these proceeedings) finds such a directional velocity
stratification arising naturally from his simulations, with
higher velocities toward the pole.

\noindent (2) 
Like our model, this hypothesis does not explain the absence of
low luminosity BALs without invoking dust.

\noindent (3) Compton scattering features will arise naturally in
all objects in this model, as in ours, if the fast wind has
sufficient column density.

\noindent (4) The fast wind has to have a large column density of
high velocity material when viewed end-on to reproduce the BAL
observations. This arises naturally in our model, but is not
obvious in this hypothesis. A large column density implies a
large mass input rate at the wind base, decaying rapidly to
larger radii (in the fast polar case; to smaller radii for the
fast equatorial case).

\noindent (5) In the fast polar case a much ($<$1\%) lower column
density is required when viewed from other directions to avoid
BALs being seen more often. A long thin BAL region is necessary,
which would have to be non-divergent to maintain column density.
This may make it difficult to cover a 10\% solid angle. An
equatorial fast wind avoids this problem.

\noindent (6) 
An equatorial fast wind has the faster material originating
further from the continuum source, which seems unlikely since
radiation pressure and Keplerian rotation both predict the
opposite. A polar fast wind avoids this problem.

The difficulties in this version of the model lie
primarily in theory: can the large column densities
needed in the fast wind be produced, in an acceptable geometry?
Neither the polar nor equatorial solution is fully appealing.

\section{Conclusions}

Because the model puts together so many aspects of quasar/AGN
phenomenology it is highly overconstrained, and so readily
tested. This is a strength of the model.  Our model has passed
quite a number of tests already, but they are not yet as
stringent as one would like.

Quite simple extensions of the model (e.g. to type~2 objects,
Risaliti, Elvis, \& Nicastro 2001) suggest that much more of the
quasar/AGN phenomenology can be incorporated with only a handful
of extra variables.  With luck then, quasars will now enter a
period of rapid development of their physics (e.g. Nicastro
2000), allowing their physical evolution to be understood, and
placing them constructively within cosmology.

\smallskip
This work was supported in part by NASA grant NAG5-6078.



\begin{references}
\reference Allen C.W. 1975, `Astrophysical Quantities' 3rd
ed. [London:Athlone], p. 198.

\reference Elvis M. 2000, ApJ, 545, 63. astro-ph/0008064

\reference Elvis M. et al. 1978, MNRAS, 183, 129

\reference Collinge M.J., et al., 2001, ApJ in press.

\reference Hamann F., Shields J., Ferland G., \& Korista K. 1995,
ApJ, 454, 688

\reference Iwasawa K. \& Taniguchi Y., 1993, ApJL, 413, L15

\reference Kaspi S., et al. 1996, ApJ, 470, 336

\reference Lawrence A., 1987, PASP, 99, 309

\reference Mathur S., Elvis M., \& Wilkes B.J., 1995, ApJ, 452, 230

\reference Mathur S., Elvis M., \& Singh K.P., 1995, ApJL, 455, L9

\reference Schild R. 1996, ApJ, 464, 125

\reference Risaliti G., Elvis M. \& Nicastro F. 2001, ApJ,
submitted

\reference Takahashi T., Inoue H., \& Dotani T., 2001

\reference Nicastro F., 2000, ApJ 530, L65





\end{references}
\end{document}